\documentclass[11pt,letter]{article}
\usepackage{graphicx}
\usepackage{amsmath}
\usepackage{amssymb}
\usepackage{cancel}
\usepackage[left=2cm,right=2cm,top=3cm,bottom=2cm]{geometry}
\usepackage{setspace}
\usepackage{float}
\usepackage[super,sort&compress,comma]{natbib} \usepackage[font={small}]{caption} 
\usepackage[labelfont=bf]{caption}
\usepackage{bm}
\usepackage[T1]{fontenc}
\usepackage{microtype}
\usepackage{subfigure}
\usepackage{chemformula}
\usepackage[version=4]{mhchem}
\usepackage{textcomp}
\usepackage{color}
\usepackage{mathtools}
\usepackage{bm}
\usepackage{gensymb}
\usepackage{esvect}
\usepackage{booktabs}
\usepackage{multirow}

\begin{document}
\title{\textbf{Learning to Evolve Structural Ensembles of Unfolded and Disordered Proteins Using Experimental Solution Data}}
\author{Oufan Zhang$^{1}$, Mojtaba Haghighatlari$^{1}$, Jie Li$^{1}$, J\~{o}ao Miguel Correia Teixeira$^{3,4}$,\\
Ashley Namini$^{3}$, Zi Hao Liu$^{3,4}$, Julie D Forman-Kay$^{3,4}$, Teresa Head-Gordon$^{1,2}$\\}
\date{}
\maketitle
\noindent
\begin{center}
$^1$Kenneth S. Pitzer Theory Center and Department of Chemistry\\
$^2$Departments of Bioengineering and Chemical and Biomolecular Engineering\\
University of California, Berkeley, CA, USA
$^3$Molecular Medicine Program, Hospital for Sick Children, Toronto, Ontario M5S 1A8, Canada\\
$^4$Department of Biochemistry, University of Toronto, Toronto, Ontario M5G 1X8, Canada

corresponding author: thg@berkeley.edu
\end{center}

\begin{abstract}
\noindent
Structural characterization of proteins with disorder requires a computational approach backed by experiment to model their diverse and dynamic structural ensembles. The selection of conformational ensembles consistent with solution experiments of disordered proteins highly depends on the initial pool of conformers, with currently available tools having issues with thorough sampling. We have developed a Generative Recurrent Neural Networks (GRNN) that is combined with a reinforcement learning (RL) step that biases the probability distributions of torsions to take advantage of experimental data types such as J-Couplings, NOEs and PREs. We show that updating the generative model parameters according to the reward feedback on the basis of the agreement between experimental data and probabilistic selection of torsions from learned distributions improves upon existing approaches that simply reweight conformers of a static structural pool for disordered proteins.  Instead the RL-GRNN learns to physically change the conformations of the underlying pool of the disordered protein to those that better agree with experiment.
\end{abstract}
\begin{figure*}
\centering
\includegraphics[width=8cm]{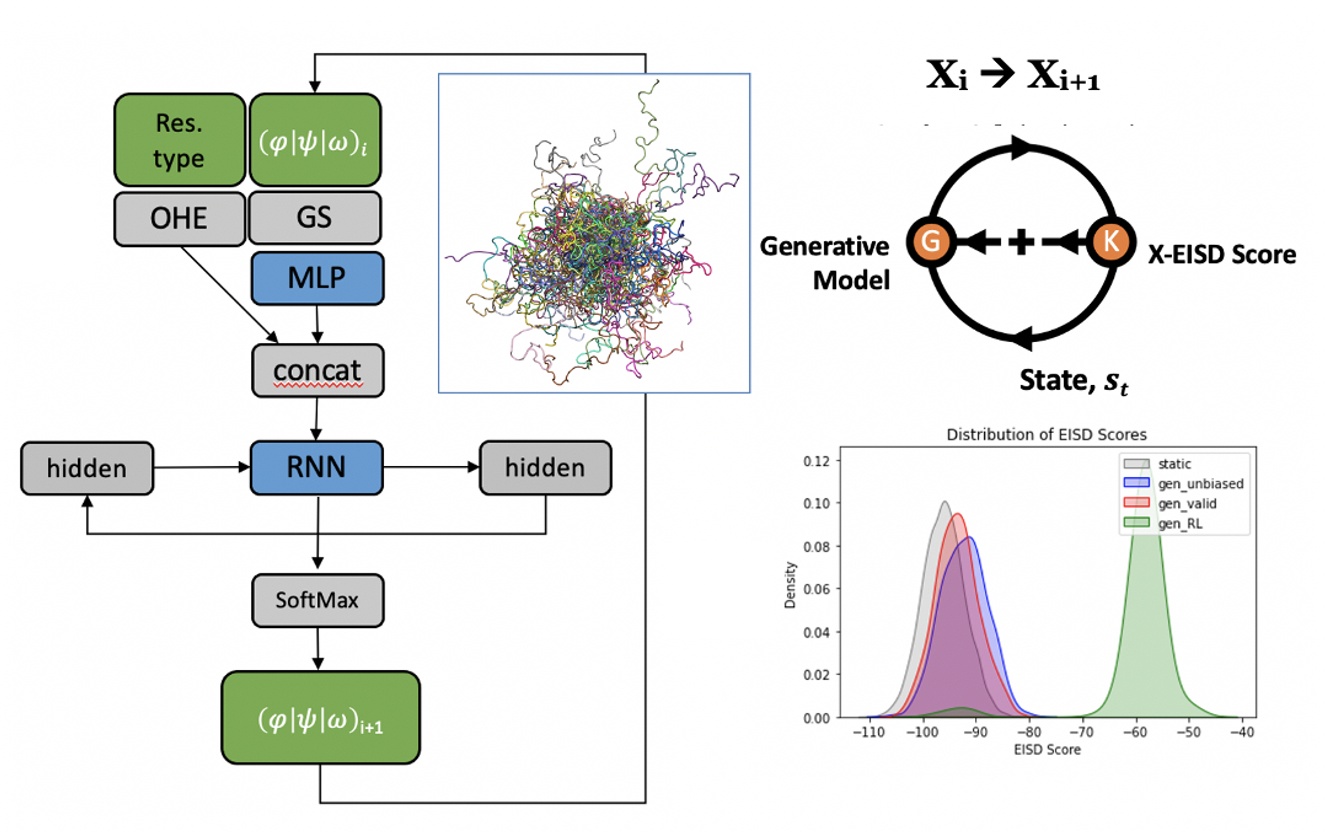}
\end{figure*}

\newpage
\section*{\fontsize{12}{12}\selectfont INTRODUCTION}
\vspace{-3mm}
For folded proteins, high resolution X-ray structures provide concrete and conceptually straightforward models, and offer a powerful connections between structure and protein function. By contrast, investigation of the functional properties of intrinsically disordered proteins (IDPs) is a significant integrative biology challenge, demanding an extensive repertoire of new experimental and computational techniques in order to create structural ensembles corresponding to the free monomer through to discrete dynamic complexes with binding partners and large-scale phase-separated states. \\

A detailed and accurate characterization of disordered protein states often requires integrating various types of biophysical experiments with computational methods. Both are needed to derive a structural ensemble that represents the conformational heterogeneity while conforming to the solution experiments that only measure ensemble and/or time averages given the dynamic nature of unfolded or intrinsically disordered proteins (IDPs). A number of approaches have been developed for generating and evaluating disordered structural ensembles that are consistent with the collective experimental restraints, including Nuclear Magnetic Resonance (NMR), small angle X-ray scattering (SAXS), single molecule fluorescence resonance energy transfer (smFRET) and any other available solution experimental measurements.\cite{Bhowmick2016,Gomes2020} \\

Creation of large structural pools of unfolded or IDP conformations can be derived from a variety of sources such as molecular dynamics (MD) simulations using a force field\cite{Ball2011,Ball2013,Ball2014}, or structural builders such as TraDES\cite{Feldman2000}, Flexible-Meccano\cite{Ozenne2012}, FastFloppyTail\cite{Ferrie2020}, and IDPConformerGenerator\cite{Teixeira2022}. To optimize agreement with experiments, most methods have typically focused on either biasing molecular simulations using experimental data as in the case of the ensemble-biased metadynamics method\cite{Allison2009}, or by selecting a collection of structures from a pre-generated pool of candidate conformers that best fits the available experimental data, such as ENSEMBLE\cite{Choy2001,Marsh2007,Marsh2009,Krzeminski2013}, Mollack\cite{Fisher2010,Fisher2011,Fisher2012}, energy-minima mapping and weighting method\cite{Huang2008,Yoon2009}, and ASTEROIDS\cite{Jensen2010,Ozenne2012,Schneider2012,Jensen2013}. \\

In recent years, Bayesian models have emerged as an ideal framework to account for the multiple and different sources of uncertainties in the IDP problem, most typically experimental and back-calculation model errors, as originally proposed by Stultz and co-workers.\cite{Fisher2010,Fisher2011,Fisher2012} These robust statistical approaches provide a confidence level in the calculated structural ensemble models given their undetermined nature and variable quality of the restraining experimental solution  data.\cite{Fisher2010,Fisher2011,Fisher2012,Hummer2015,Brookes2016,Bonomi2016,Bonomi2017,Kofinger2019,Lincoff2020,Kofinger2019inferring,Ahmed2021,Bottaro2020} Among some of the most visible developments are maximum parsimony inspired methods exemplified by Bayesian weighing (BW) method\cite{Fisher2010}, and maximum entropy inspired techniques represented by the Bayesian ensemble refinement method\cite{Hummer2015}, Metainference\cite{Bonomi2016,Ahmed2021}, Bayesian/Maximum entropy (BME)\cite{Bottaro2018,Bottaro2020} and the Bayesian inference of ensembles (BioEn) method\cite{Kofinger2019inferring,Kofinger2019}. The Head-Gordon lab developed the extended Experimental Inferential Structure Determination (X-EISD) method that treats experimental and model errors as Gaussian random variables, and can use their joint probabilities in a Monte Carlo sampling or maximization procedure for refining the computational ensembles.\cite{Brookes2016,Lincoff2020} \\

But in order for these Bayesian approaches to be successful requires the underlying structural pool to cover a representative conformational space, such that the most important conformers can be weighted more heavily than more irrelevant conformations for the optimization to be effective. However the "putative" disordered ensemble may not contain a relevant pool of structures. For example, some structural builder approaches\cite{Feldman2000,Ozenne2012,Ferrie2020} can generate structures that are unphysical, with large steric clashes and lacking Boltzmann weighting. While MD-generated ensembles do contain energetically weighted states, they have clear structural biases towards overly compact states using many of the current force fields, and thus are poor descriptions for disordered protein states.\cite{Liu2021} While new IDP-specific force fields have been introduced\cite{Best2010,Piana2015,Huang2017}, in some cases they no longer describe folded states\cite{Robustelli2018} and/or tend to become too unstructured and featureless to be consistent with the solution data\cite{Fluitt2015}, and therefore can result in underlying biases in the resulting ensemble. \\

Various deep learning models, most notably AlphaFold2\cite{Jumper2021} and RoseTTAFold\cite{Baek2021}, have made stunning breakthroughs in producing target structures of monomeric folded proteins of quality similar to experimental structures.\cite{Torrisi2020,Masrati2021,Wang2017,Schaarschmidt2018,Adhikari2018,Alquraishi2021,Alquraishi2019} Other examples are deep convolutional neural networks that predict structures as distance maps\cite{Wang2017,Schaarschmidt2018,Adhikari2018,Alquraishi2021}, and in leveraging natural language processing to encode protein sequences using recurrent neural networks (RNNs)\cite{Alquraishi2019,King2021}. Advances in structure prediction and generation for folded proteins foreshadow an exciting frontier of applying machine learning methods in the integrative modelling of IDP ensembles.\cite{Lindorff2021,Ramanathan2021} To create a diverse and representative protein structural space, the machine learning field has also seen an emergence of generative neural networks\cite{Hoseini2021}, predominantly employing variational autoencoders (VAEs) and generative adversarial networks (GANs) to learn from native protein databases to propose structural variants of folded states\cite{Guo2021, Rahman2021}, or to learn from MD to provide a less computationally expensive alternative for conformational sampling\cite{Degiacomi2019,Moritsugu2021}. Recently, Gupta \textit{et al.} used a VAE to compress MD generated conformers for the disordered $\alpha\beta$40 and ChiZ proteins to a low-dimensional latent space, which are sampled to reconstruct conformers and subsequently validated against NMR chemical shifts and SAXS data\cite{Gupta2022}. Janson \textit{et al.} also proposed a GAN model called idpGAN to learn course-grained simulations of disordered proteins\cite{Janson2022}.\\

Here we introduce a machine learning approach that learns the probability of the next residue torsions $X_{i+1}=\ [\phi_{i+1},\psi_{i+1},\omega _{i+1}, \chi_{i+1}]$ from the previous residue in the sequence $X_i$ using a generative recurrent neural network (GRNN) model to build new conformational states of a disordered protein ensemble. This work is distinguished by a further reinforcement learning (RL) step that biases the probability distributions of torsions of the GRNN to take advantage of experimental data types such as three-bond J-couplings, nuclear Overhauser effects (NOEs) and paramagnetic resonance enhancements (PREs) from NMR spectroscopy. The resulting RL-GRNN machine learning model, which we call DynamICE (dynamic IDP creator with experimental restraints), learns to physically change the conformations of the underlying pool to those that better agree with solution experiments, using the X-EISD Bayesian model and enforcing realistic energetic states through a Lennard-Jones potential. We show that updating the DynamICE model parameters according to the reward feedback on the basis of the agreement between structures and data improves upon existing approaches that simply reweight static structural pools for disordered proteins.\cite{Hummer2015,Lincoff2020,Bottaro2020} \\

\section*{\fontsize{12}{12}\selectfont METHODS}
\begin{figure*}[!ht]
\centering
\includegraphics[width=17cm]{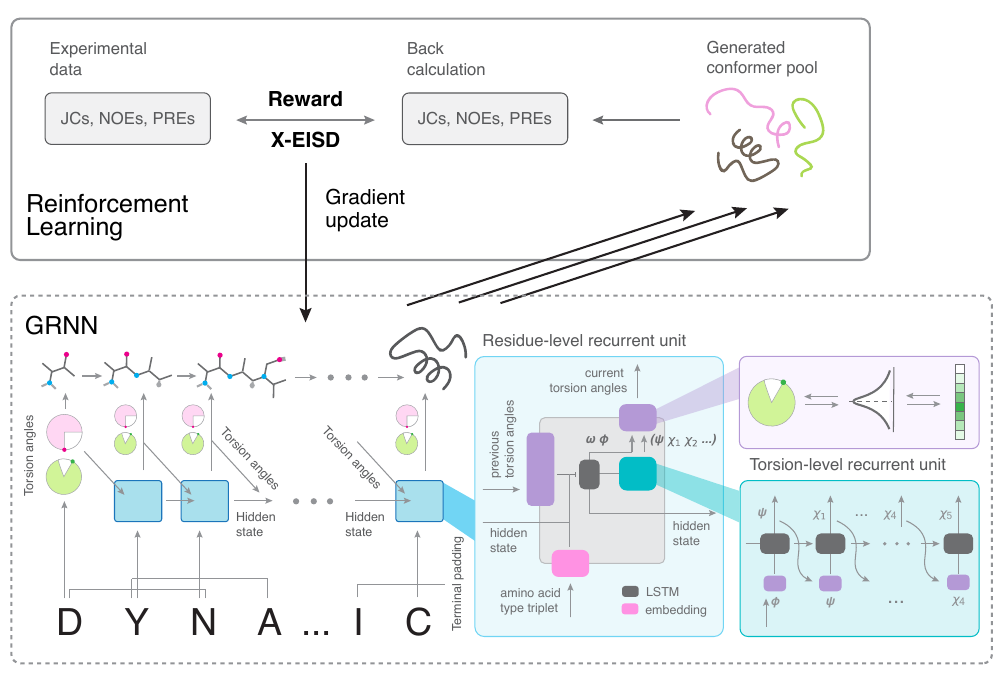}
\caption{\textbf{Schematic of the design of the DynamICE GRNN and its interplay with reward evaluation using X-EISD to evolve new conformer generation in a reinforcement learning workflow.}  Bottom: The recurrent unit (sky blue) takes at a time a triplet of the adjacent residue types, torsion angles of the previous residue, and the previous hidden state to compute an internal state to generate torsion angles, which are sequentially translated to Cartesian coordinates to generate conformers. Top: the generated conformer pools are evaluated by their agreement with the experimental data to formulate a feedback to the GRNN. Bottom-rightmost inset (teal) shows the recurrent unit that handles torsion angles generation within a residue step. Middle-rightmost inset (purple): the two-way representation from a torsion angle to a Gaussian smeared probability vector as network input, and sampling of a predicted probability vector to a torsion angle.}\label{fig:network}
\end{figure*}

\vspace{-3mm}
Figure \ref{fig:network} illustrates the DynamICE reinforcement learning cycle, where the generative recurrent neural network (GRNN) generates conformers and interacts with X-EISD to evolve new conformers in better agreement with the experimental data. In this section we describe the protein structure representation used, the design of the GRNN, and the reinforcement learning workflow in detail. Some more technical details of training are further described in the Supplementary Information.\\

\noindent
\textit{Protein conformer representation}. Assuming ideal bond lengths and bond angles, a protein conformer can be represented by a sequence of backbone and sidechain torsion angles for each residue j ($\omega_j$, $\phi_j$, $\psi_j$, $\chi_j1$, $\chi_j2$, ..., $\chi_j5$). By parameterizing protein structures in the torsional space, the generative model covers diverse conformations in a reduced dimension while preserving local chemical connectivity. The torsional space is discretized into 2 degree bins over the range of $[-\pi, \pi)$ such that each torsion angle is represented by a vector of size 180 with elements corresponding to the relative probability of finding the angle at each angle bin. Torsion angles are smoothed out as Gaussian distributions with a 1 degree standard deviation to allow for flexibility, and periodic boundaries are enforced.  \\

\noindent
\textit{Generative recurrent neural network architecture and conformation generation}
RNNs are designed to handle sequential information by determining the current outputs from past information along with the current inputs. In this work, we use an advanced  multi-layer long short-term memory (LSTM) network\cite{Hochreiter1997}, to predict a distribution of an accessible angle range of the torsion angles in the current residue given those of the last residue and its associated hidden state. LSTMs can preserve long-term memory while ignoring certain short-term inputs in a dedicated mechanism.\cite{Hochreiter1997}  The basic LSTM cell contains two internal states, the hidden state $h_t$ and the cell state $c_t$, and can be described through the following set of equations:
\begin{align}
    i_t&=\sigma(W^ix_t+U^ih_{t-1}) \\
    f_t&=\sigma(W^fx_t+U^fh_{t-1})  \\
    o_t&=\sigma(W^ox_t+U^oh_{t-1}) \\
    \Tilde{c}_t&=\tanh{(W^cx_t+U^ch_{t-1})} \\
    c_t&=i_t\odot\Tilde{c}_t+f_t\odot c_{t-1} \\
    h_t&=o_t\odot\tanh{c_t}
\end{align}
where $[W^i,W^f,W^o,W^c,U^i,U^f,U^o,U^c]$ are the trainable parameters of the model, $x_t$ is the input to the cell at the current timestep, $\Tilde{c}_t$ contains the information to be added to the cell state, and $i_t,f_t,o_t$ represent the update gate, forget gate and output gate respectively, which are numbers between $(0,1)$ that controls how much information should be updated, discarded or retrieved from the cell state. $\sigma$ denotes the sigmoid function, and $\odot$ represents element-wise multiplication. \\

The recurrent units inherently formulate a conditional probability between individual torsion angles at the local level that are chained to create a global representation of the entire chain. The 8 torsion angle vectors representing the backbone and sidechain torsion angles in a residue are concatenated with a 64 length embedding layer that encodes the amino acid type of a triplet of the previous, current and subsequent residue. We use alanine as a terminal padding for the last amino acid type triplet. Together they are transformed through a 2-layer fully connected multi-layer perceptrons (MLP) with a Rectified Linear unit (ReLU) activation for each layer. Torsions of residues with less than 5 sidechain angles are padded with zero.\\

The generative model contains 2 recurrent units, one for recursion between residues and one for recursion between torsion angles within a residue. This design allows the model to capture correlations of torsion angles between residues as well as correlations between torsion angles within a residue. In the residue-level recurrent unit, the multi-layer perceptron (MLP) outputs are passed to a RNN cell connected to 2 linear layers corresponding to the $\omega$ and $\phi$ torsion angles. The torsion-level recurrent unit is enclosed inside the residue-level unit and iterates through the rest of the torsion angles ($\psi$, $\chi_1$, $\chi_2$, ...) using the generated $\phi$ angle. Along with the torsion angle vectors and the MLP outputs, an one-hot encoding of torsion angle types is passed to a RNN cell connected to a linear layer. Each linear layer uses a softmax activation to transform the output into a vector that represents the probability distribution of a torsion angle. The residue-level RNN cell contains 2 stacked LSTMs with hidden size of 200 and dropouts of 0.1, while the torsion-level RNN cell uses 1 LSTM with the same hidden size and dropout configurations. The generative model is implemented using PyTorch. We describe the details of the pre-training procedure of the generative model in the Supplementary Information.\\

To initiate the generation of a new protein conformer, a set of torsion angles of the first residue along with its protein sequence is provided to the generative model. The model repeatedly takes the torsion angles of the current residue to generate the probability distributions from which the torsion angles of the next residue are sampled until it reaches the last residue. The torsion angles are translated to Cartesian coordinates to generate a conformer. A Lennard-Jones potential is computed using Amber14SB parameters\cite{Hornak2006} with a user-definable threshold to reject severe clashes at each residue iteration during the conformer building process. The building and validation of conformers are supported by a conformer generator module adapted from IDPConformerGenerator\cite{Teixeira2022}.\\

\noindent
\textit{Reinforcement learning procedure}
The goal of RL is to learn an optimal strategy of actions that maximizes the expected return, which can be approximated as the sum of rewards $r_\Theta$ with network parameter $\Theta$ through sampling the state-action space $s_T$,
\begin{align}
    J(\Theta) &= \mathbb{E}_{s_T\sim p(s_T)}[r_{\Theta}(s_T)]\\                      &\approx\sum_{s_T} r_{\Theta}(s_T)\\
              &= - \sum_{s_T} (V(s_T, \Theta) - \hat{V})^2. \label{eqn:rlloss}
\end{align}
This is analogous to minimizing the loss between the back calculation $V(s_T, \Theta)$ of an ensemble of sampled structures (trajectories of torsion angles) and the target experimental observables $\hat{V}$.  
To embrace the uncertainties $\sigma_{exp}$ associated with these experimental data, we also devise a 'flat-bottom' loss by only performing gradient updates on the terms of which the back calculations are outside of the experimental uncertainty ranges. 
\begin{align}
    J(\Theta) &= - \sum_{V(s_T, \Theta) \notin [\hat{V}-\sigma_{exp}, \hat{V}+\sigma_{exp}]} (V(s_T, \Theta) - \hat{V})^2. \label{eqn:rlfbloss}
\end{align}

We train models that are biased with J-couplings (JCs), nuclear Overhauser effects (NOEs) and paramagnetic relaxation enhancements (PREs). J-couplings (JCs) are defined by the backbone $\phi$ torsion angle $H_N-N-C_\alpha-H_\alpha$ and the ensemble average are back calculated using the Karplus equation\cite{Karplus1963} as,
\begin{align}
    V(\phi) = \langle A\cos{(\phi - \phi_0)}^{2}+B\cos{(\phi - \phi_0)}+C\rangle,
\end{align}
where $\phi_0$ is a reference state offset of 60 degrees, and $A$, $B$, and $C$ are back calculation parameters sampled as random Gaussian variables\cite{Lincoff2020} with mean and standard deviation values provided in the work of Vuister and Bax\cite{Vuister1993}. NOEs and PREs back calculations are modeled as the ensemble averaged distance $D$ of $N$ structures using the ENSEMBLE approach\cite{Krzeminski2013,Marsh2007,Marsh2009},
\begin{align}
    D = (\frac{\sum^{N}_{i=1} {d_i}^{-6}}{N})^{-1/6}.
\end{align}
For joint optimization with multiple data types, the total reward function sums up the reward for each data type according to Eqn.\ref{eqn:rlfbloss} with a weight hyperparameter. We describe the details of the RL training procedure in the Supplementary Information.

To keep the gradient information of the back calculations generated from the sampled torsion angles, we utilize Gumbel-Softmax\cite{Jang2016} as a differentiable reparameterization trick that allows sampling from a categorical distribution of $i$ classes during the forward pass of a neural network. The sample vector $y_i$ from the generated torsion distribution with probabilities $p_i$ is expressed as
\begin{align}
    y_i = \frac{exp((log(p_i)+g_i)/\lambda)}{\sum_{i} exp((log(p_i)+g_i)/\lambda)},
\end{align}
where $g_i$ denotes noise generated from a Gumbel distribution, and the softmax function is taken over the reparameterized distribution with a temperature hyperparameter $\lambda$. We use an annealing schedule that starts from 1 and gradually decreases the temperature by an order of 0.98 for each training iteration. This annealing process balances between model accuracy and variance associated with temperature: the models are trained robustly with low variance at high temperature initially, and as the model parameters began to converge, the temperature lowering ensures accuracy without causing significant instability\cite{Jang2016}. This recast of a stochastic generation process allows the model to trace the rewards based on distance type restraints to specific torsion angles through an internal to Cartesian coordinates conversion (see Supplementary Information), thereby overcoming difficulties a generative model defined in torsional space being less sensitive to tertiary contact restraints such as NOEs and PREs as compared to local and angular restraints such as backbone J-couplings. 

The best model is selected based on the X-EISD score of the generated structures during the validation steps. The use of a Bayesian model for validation furnishes the RL generative model with a better probabilistic interpretation of disordered protein ensembles by modulating different sources of uncertainties in the experimental data types. We briefly summarize the details of the X-EISD approach in the Supplementary Information, and refer readers to references \cite{Brookes2016,Lincoff2020}.

\noindent
\textit{Data and code availability}
The DynamICE package has been implemented as a publicly accessible python package at https://github.com/THGLab/DynamICE for the reproducibility of future studies.

\section*{\fontsize{12}{12}\selectfont RESULTS}
We apply our DynamICE learning approach to two protein cases: the unfolded state of the \textit{Drosophila} DrkN-terminal SH3 domain (uDrkN-SH3) and the $\alpha$-synuclein ($\alpha$-Syn) IDP to demonstrate its ability to evolve new conformers driven by a better agreement with solution experimental data. uDrkN-SH3, which exists in approximately 1:1 equilibrium between folded and unfolded states under non-denaturing conditions, is a popular test case with abundant experimental data made available for ensemble reweighting programs for disordered protein\cite{Marsh2007}. $\alpha$-Syn is an IDP linked to a set of neurodegenerative diseases\cite{Goedert2001}, and also has ample experimental data. \\

Both initial conformer pools for uDrkN-SH3 and $\alpha$-Syn are generated using IDPConformerGenerator.\cite{Teixeira2022} IDPConformerGenerator is a flexible software platform that can be used to create conformers drawing on torsion angles from any secondary structure combination. In this work we use it to randomly sample only loop and extended state torsion angles to make a conformer pool lacking helix for uDrkN-SH3, since the uDrkN-SH3 protein is known to have local regions of helical structure, in order to generate a clear test case with a starting pool missing relevant conformers. Correspondingly for $\alpha$-Syn, we randomly sampled only loop and helical state torsion angles to make a conformer pool without extended conformations, although the $\alpha$-Syn protein is largely known to be featureless in secondary structure signatures. These are intended to be challenging cases, i.e., the underlying conformational pools are poor "start states" for a reweighting algorithm. But as a control, we also start with randomly sampled torsion angles comprising loop and extended states for $\alpha$-Syn which should favor the reweighting algorithm, i.e., a  starting pool in which relevant conformations are (fortuitously) present, to see how the DynamICE compares for this special case. \\

\begin{figure*}[ht!]
\centering
\includegraphics[width=17cm]{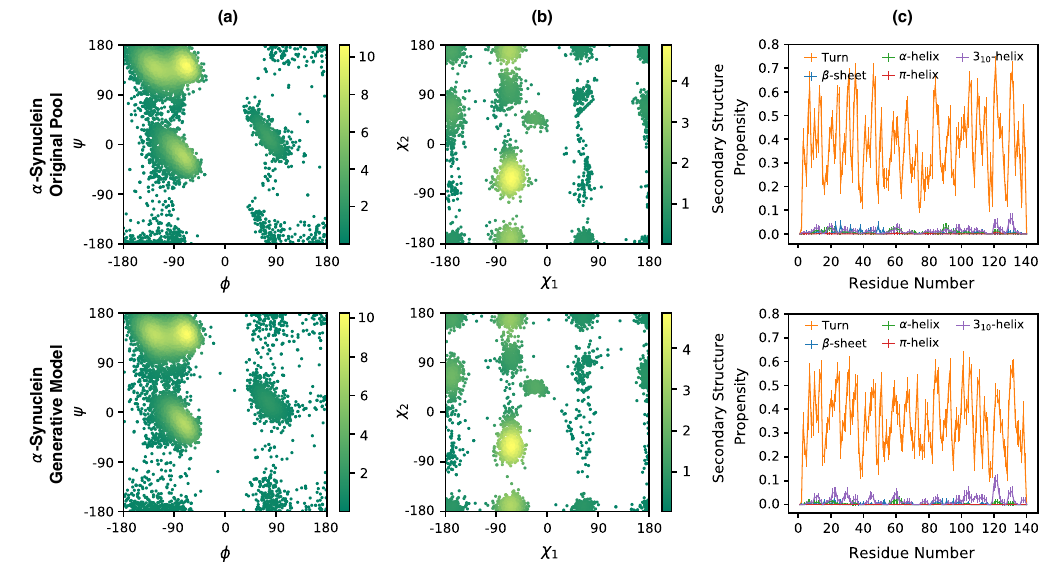}
\caption{\textbf{Properties of ensembles for the $\alpha$-Syn IDP from the original pool and from the generative model.} a) Ramachandran plots displaying the backbone torsion angle distributions and b) Histograms displaying the $\chi_1-\chi_2$ distributions from 100 structures of the training data (top) and generative model (bottom). Density values are scaled by 1e-05. c). Secondary structure propensities per residue among 50 independently drawn ensembles of 100 structures. Error bars are shown as $\pm$ 1 standard deviation. Corresponding plot for uDrkN-SH3 is shown in Fig. S1}
\label{fig:asyn_pretrain}
\end{figure*}

We begin with the unbiased generative model that learns the torsional statistics of backbone and side chains of the given protein sequence given the respective starting conformational pools from IDPConformerGenerator. The performance of the generative model in this "pre-training" phase are evaluated by a number of metrics: how closely the Ramachandran plot for backbone and sidechain torsions from the generative model matches the corresponding Ramachandran plot of the original training conformer pools, and the agreement with the percentage of secondary (local) structure per residue of all the major secondary structure categories. Fig. \ref{fig:asyn_pretrain} displays the results for the $\alpha$-Syn IDP with a starting pool of loops and helices, and similar plots are made for the loop/extended states used for uDrkN-SH3 (Fig. S1). Table S1 shows that the underlying structural differences of the original pool and generative ensembles are minimal in terms of global shape characteristics such as the radius of gyration $R_g$, end-to-end distance $R_{ee}$, and asphericity $\delta^\ast$ which measures the anisotropy of the ensemble. Finally, Tables S1 also demonstrates that the generative models and their respective original conformer pools also score similarly on various experimental data types, providing additional evidences that the unbiased generative models are robust. \\

After the pre-training step we bias the GRNN toward generating new underlying conformers such that the resulting ensembles better agree with the measured experimental data. This requires a RL step that uses a reward based on minimizing the error between back calculations from the structural ensemble of observables that connect to available solution data. Here we report models trained using the 'flat-bottom' loss function Eqn.\ref{eqn:rlfbloss} which is designed to interpret the experimental errors and uncertainties. For the uDrkN-SH3, we perform an RL-GRNN optimization with J-couplings (JCs) and Nuclear Overhauser Enhancements (NOEs). We chose these two data types as our previous study with X-EISD has shown that dual reweighting optimization of local data such as JCs and long-ranged restraints such as NOEs can yield ensembles that simultaneously improve other data types\cite{Lincoff2020}. For $\alpha$-Syn, we train DynamICE by jointly optimizing JCs and PREs. PREs are a more difficult class of experimental data as they tend to be biased toward longer range contacts between different parts of the sequence compared to NOEs, which requires the model to cooperatively change torsion angles of a large number of residues to meet a contact restraint. To best demonstrate the ability of the RL-GRNN despite the torsional representation limitation, we optimize both on a subset of PRE data that only contains contacts equal to to or less than 10 residues apart as well as the full set of PRE data.

Table 1 shows that the RL-GRNN optimization reduces the RMSD of NOE data with respect to the unoptimized ensembles of uDrkN-SH3, and yields similar RMSDs for the NOEs using reweighting. However DynamICE shows a significant RMSD improvement for J-couplings by changing the conformers of the underlying ensemble. As a result of these new members an independent validation shows that global shape metrics such as PREs, smFRET, and SAXS are also improved although these additional experimental data types were not optimized directly. 

\begin{table}[h!]
\centering
\caption{\textbf{Evaluation of the RL-GRNN and reweighted ensemble optimizations for uDrkN-SH3 and $\alpha$-Syn with experimental data types and geometric measures.} The RL-GRNN models are trained with the 'flat-bottom' loss function Eqn.\ref{eqn:rlfbloss}. The experimental data RMSDs including J-couplings (JCs), nuclear overhauser effect (NOEs), paramagnetic relaxation enhancement (PREs), single molecule FRET (smFRET), chemical shifts (CS), small angle X-ray scattering (SAXS). The experimental (exp) for CSs ($\sigma_{exp}$=0.03-0.3 ppm); JCs ($\sigma_{exp}$=0.5); NOEs ($\sigma_{exp}$=5.0 Å); PREs ($\sigma_{exp}$=5.0 Å); smFRET <E> ($\sigma_{exp}$=0.02); SAXS ($\sigma_{exp}$=0.0008-0.002). For $\alpha$-Syn, we utilize all PRE distances that are less than 10 residues apart. Global metrics of the ensembles includes radius of gyration $R_g$, end-to-end distance $R_{ee}$ and ensemble asphericity $\delta^\ast$ (which measures the anisotropy of the structures ranging from 0 (sphere) to 1 (rod)). All values are reported in terms of mean and standard deviation (in parenthesis) over 50 ensembles of 100 structures each.}
\resizebox{\textwidth}{!}{\begin{tabular}{lccccccccc}
 & JC & NOE & PRE & smFRET $\langle E\rangle$ & CS & SAXS & $R_g$ & $R_{ee}$ & $\delta^\ast$\\
 & (Hz) & (\AA) & (\AA) & & (ppm) & (Intensity) & (\AA) & (\AA) & \\
\hline
\multicolumn{10}{l}{\textbf{uDrkN-SH3 ensembles UNOPTIMIZED (loop/extended) and OPTIMIZED with JCs and NOEs}} \\
\hline
Generative model  & 1.440 & 6.343 & 7.711 & 0.228 & 0.495 & 0.007 & 23.16 & 55.51 & 0.431 \\
& (0.028) & (0.429) & (1.193) & (0.032) & (0.007) & (0.000) & (4.81) & (21.21) & (0.202) \\
Reweight & 1.398 & 5.208 & 7.213 & 0.208 & 0.493 & 0.007 & 22.35 & 52.95 & 0.421 \\
    & (0.017) & (0.365) & (1.381) & (0.027) & (0.009) & (0.000) & (4.33) & (19.26) & (0.192) \\ 
RL-GRNN model & 0.693 & 5.242 & 6.346 & 0.119 & 0.478 & 0.004 & 20.28 & 48.58 & 0.401 \\
    & (0.033) & (0.410) & (1.073) & (0.035) & (0.010) & (0.000) & (3.68) & (17.52) & (0.193) \\
\hline
\multicolumn{10}{l}{\textbf{$\alpha$-Syn ensembles UNOPTIMIZED (helix/loop) and OPTIMIZED with JCs and PREs (all data)}} \\
\hline
Generative model & 0.622 & & 9.923 & 0.103 & 0.612 & 0.017 & 33.99 & 78.66 & 0.426 \\
& (0.032) & & (0.351) & (0.004) & (0.019) & (0.002) & (7.61) & (33.80) & (0.196) \\ 
Reweight & 0.528 & & 6.372 & 0.112 & 0.638 & 0.014 & 35.09 & 83.48 & 0.444 \\
 & (0.048) & & (0.158) & (0.004) & (0.026) & (0.001) & (7.30) & (32.45) & (0.201) \\ 
RL-GRNN model & 0.524 & & 8.992 & 0.145 & 0.566 & 0.025 & 43.81 & 101.25 & 0.441 \\
 & (0.017) & & (0.355) & (0.005) & (0.002) & (0.002) & (9.44) & (39.38) & (0.198) \\ 
 \hline
\multicolumn{10}{l}{\textbf{$\alpha$-Syn ensembles UNOPTIMIZED (loop/extended) and OPTIMIZED with JCs and PREs (all data)}} \\
\hline
Generative model & 0.704 & & 10.088 & 0.108 & 0.558 & 0.013 & 37.18 & 84.73 & 0.443 \\
 & (0.022) & & (0.395) & (0.005) & (0.003) & (0.001) & (7.90) & (34.32) & (0.187) \\ 
Reweight & 0.622 & & 6.200 & 0.119 & 0.550 & 0.014 & 38.49 & 87.89 & 0.432 \\
 & (0.015) & & (0.175) & (0.005) & (0.004) & (0.001) & (8.63) & (35.17) & (0.202) \\ 
RL-GRNN model & 0.655 & & 9.365 & 0.133 & 0.588 & 0.026 & 42.37 & 98.80 & 0.426 \\
 & (0.009) & & (0.309) & (0.010) & (0.003) & (0.002) & (9.91) & (40.57) & (0.192) \\ 
\hline
\bottomrule
\end{tabular}}
\label{table:drk_rlresult}
\end{table}

\begin{figure*}[!ht]
\centering
\includegraphics[width=17cm]{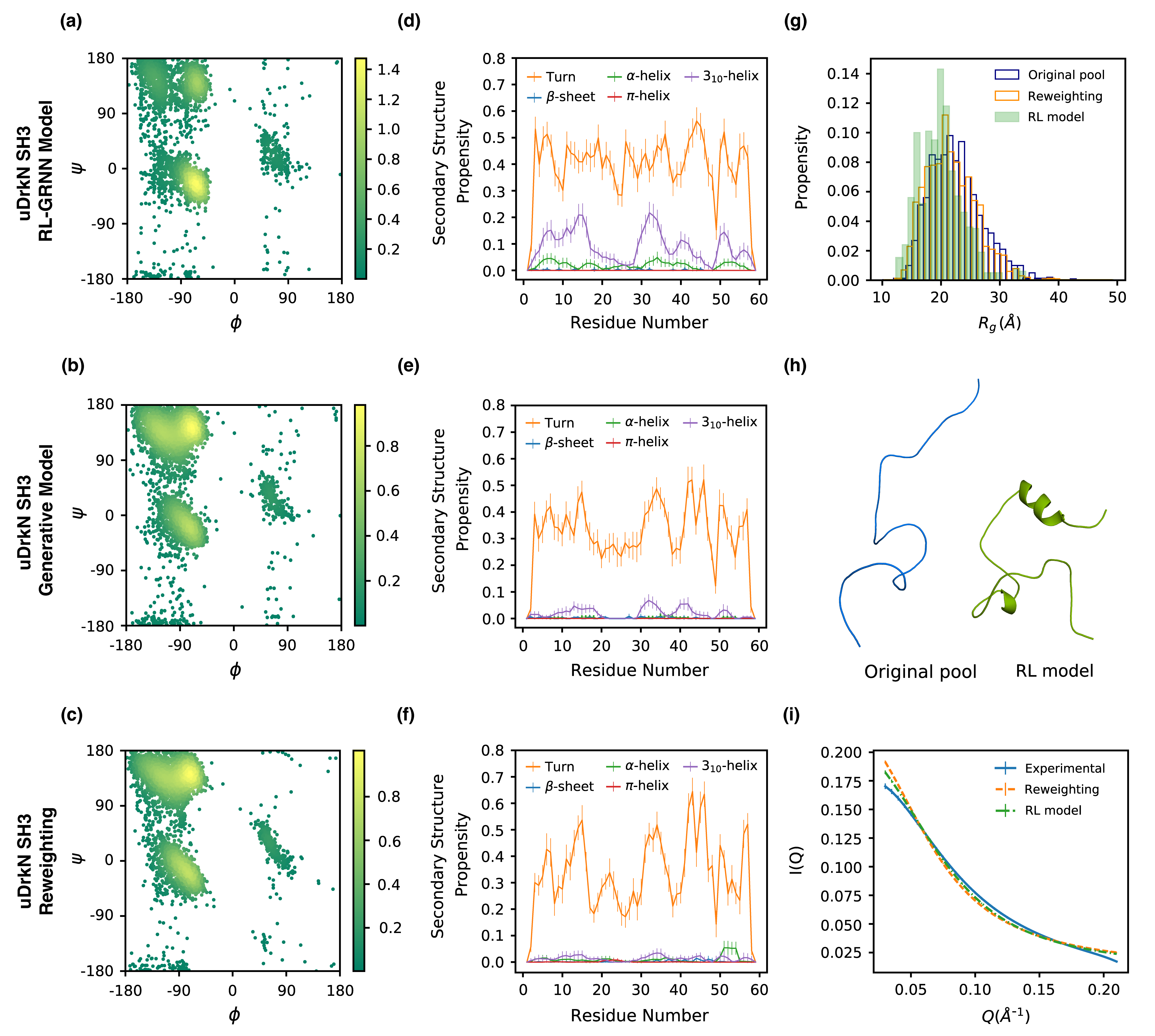}
\caption{\textbf{Properties of the uDrkN-SH3 domain unbiased ensemble and generated by the DynamICE RL-GRNN model compared with reweighting optimization using JCs and NOEs.} Ramachandran plots displaying the backbone torsion angle distributions from the (a) RL-GRNN, (b)unbiased generative model, (c) reweighting optimization. Density values are scaled by 1e-04. Secondary structure propensities per residue of the (d) RL-GRNN, (e) unbiased generative  model, (f) reweighting optimization. (g) Comparison of radius of gyration distributions before and after optimization with reweighting and RL-GRNN. (h) example of conformers from the uDrkN-SH3 original pool and RL-GRNN model, (i) SAXS intensity curves for RL-GRNN and reweighting optimized ensembles compared with the experimental data. Statistical errors from 50 independently drawn ensembles of 100 structures. Error bars are shown as $\pm$ 1 standard deviation.}
\label{fig:drk_rl}
\end{figure*}

Figure \ref{fig:drk_rl} illustrates that these improvements in the optimized and validated metrics for uDrkN-SH3 arise because the backbone torsion angles shift towards the helical region ($\phi$ = -80$^\circ$ to -50$^\circ$ and $\psi$ = -25$^\circ$ to -60$^\circ$) after RL-GRNN optimization (Fig. \ref{fig:drk_rl}a), leading to a substantial increase in the percentage of helical content from nearly zero to around 10\% - 30\% at residues 10-20 and 30-45 as seen in Fig. \ref{fig:drk_rl}d. The favored helical regions are similar to those in the optimized ensembles of uDrkN-SH3 in previous studies\cite{Marsh2009,Lincoff2020}, and are supported by JCs and the NOE data which include a number of $i$ to $i+3$ or $i+4$ contacts around residue 15-20 and 30-40. By contrast, the reweighting optimization barely changes the torsion angle profiles from the unbiased pools (Figs. \ref{fig:drk_rl}b,c) nor is there a shift in the secondary structure assignments (Figs. \ref{fig:drk_rl}e,f) due to a lack of relevant conformers in the initial pool to further refine ensembles using the JCs and NOEs data. While reweighting optimization yield ensembles that slightly shift toward more compact and globular-like conformers as measured by $\langle R_g\rangle$, $\langle R_{ee}\rangle$ and $\langle \delta^\ast\rangle$, the RL-GRNN model shows a more pronounced shift in $R_g$ (Fig. \ref{fig:drk_rl}g) and $R_{ee}$ distributions (Table 1) to even more compact disordered states. These new sub-population of helical structures (Fig. \ref{fig:drk_rl}h) and more compact conformers generated by DynamICE, which are not available in the original pool used in the reweighting scheme, are responsible for the better agreement with the overall SAXS intensity profile (Fig. \ref{fig:drk_rl}i). Thus by generating physically different conformers the RL-GRNN model directly overcomes the deficiencies of the static initial ensemble.

\begin{figure*}[!ht]
\centering
\includegraphics[width=17cm]{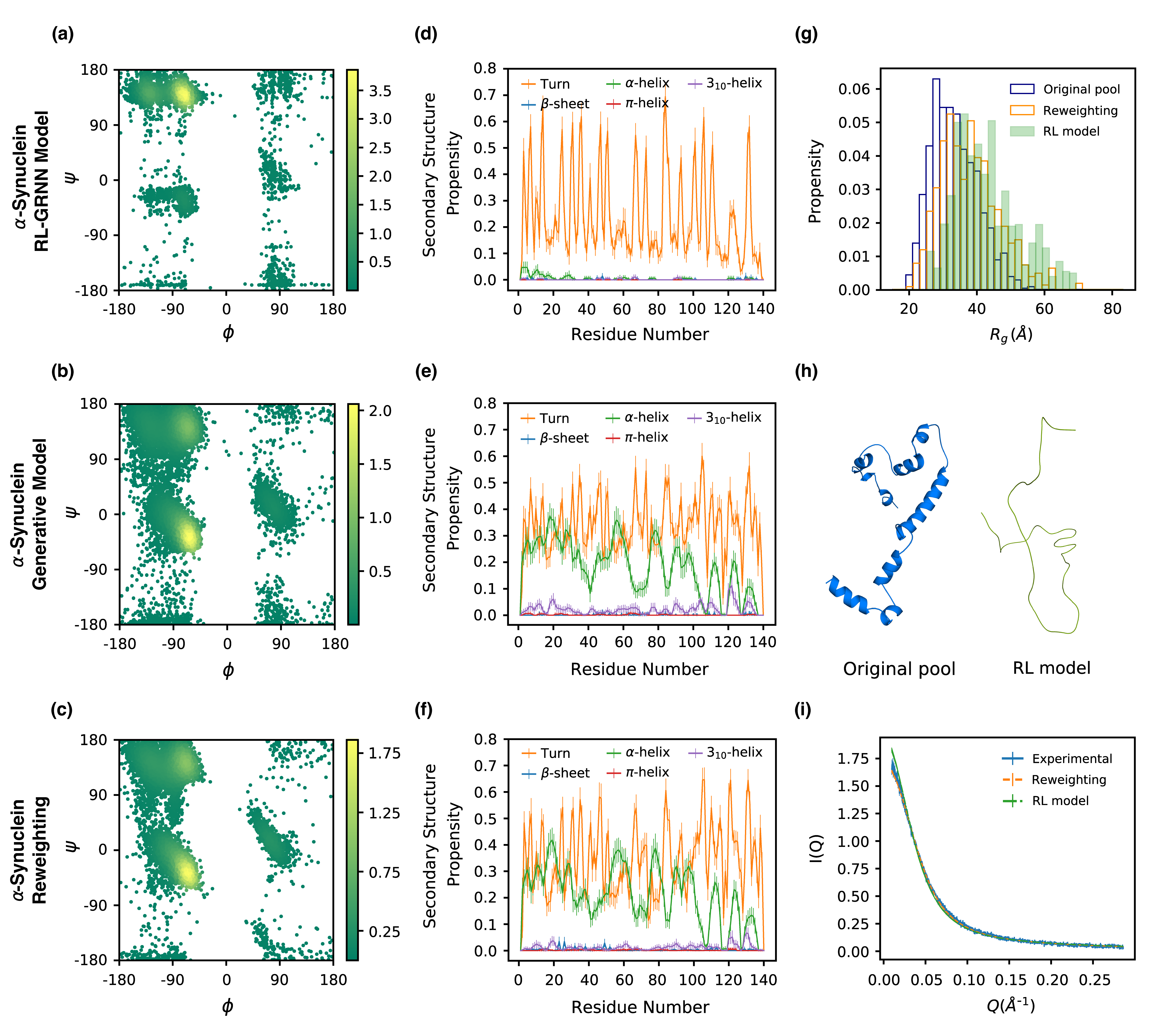}
\caption{\textbf{Properties of the $\alpha$-Syn unbiased ensemble and generated by the DynamICE RL-GRNN model compared with reweighting optimization using JC's and PREs.} Ramachandran plots displaying the backbone torsion angle distributions from the (a) RL-GRNN, (b)unbiased generative model, (c) reweighting optimization. Density values are scaled by 1e-04. Secondary structure propensities per residue of the (d) RL-GRNN, (e) unbiased generative  model, (f) reweighting optimization. (g) Comparison of radius of gyration distributions before and after optimization with reweighting optimization and RL-GRNN. (h) Example of conformers from the $\alpha$-Syn original pool and RL-GRNN model. (i) SAXS intensity curves for RL-GRNN and reweighting optimized ensembles compared with the experimental data. Statistical errors from 50 independently drawn ensembles of 100 structures. Error bars are shown as $\pm$ 1 standard deviation.}
\label{fig:asyn_rl}
\end{figure*}

The RL-GRNN results for the $\alpha$-Syn IDP starting from an unbiased pool containing loops and helices are shown in Fig. 4 and Table 1. As in the case for uDrkN-SH3, both the reweighting and RL-GRNN models achieve improvements in J-Couplings and PRE data types compared with the unbiased generative model for $\alpha$-Syn (Table \ref{table:drk_rlresult}). Given the uncertainties in PRE distances of 5$\AA$ the reweighting and RL-GRNN RMSDs are not distinguishable. But unlike the case of uDrkN-SH3 in which DynamICE drove to more compact ensembles, the combination of J-coupling and PRE data drives the backbone torsion angles towards the polyproline-II region ($\phi$ = -90$^\circ$ to -25$^\circ$ and $\psi$ = 120$^\circ$ to 150$^\circ$) after RL-GRNN optimization (Fig. \ref{fig:asyn_rl}a) compared to the unbiased and reweighted ensembles (Figs. \ref{fig:asyn_rl}b,c). In particular we see that the DynamICE optimization has largely eliminated helical torsions when compared to the unbiased and reweighted ensembles (\ref{fig:asyn_rl}d,e,f), as these conformational states are not unambiguously supported by the JC and PRE experimental data. Since the RL-GRNN model introduces more extended conformations \ref{fig:asyn_rl}(h)), it noticeably shifts the $R_g$ (Fig. \ref{fig:asyn_rl}(g)) and $R_{ee}$ (Table \ref{table:drk_rlresult}) distributions in comparison with the unoptimized and reweighted ensembles, and exhibits good agreement with the SAXS data ((Fig. \ref{fig:asyn_rl}(i))).  Since no qualitative differences are found when starting from the loop/extended vs helix/loop pools, we believe that the limited experimental data of J-couplings and NOEs as formulated can't distinguish between the two qualitatively different ensembles optimized by the RL-GRNN and reweighting approaches, and both are reasonably validated by the other data types. In fact, Table S2 and Figure S3 shows that if we restrict the PREs to be short-ranged in sequence separation (labels less than 10 amino acids apart) the support for more extended states is still strong although long-range PRE data is missing. 

\section*{\fontsize{12}{12}\selectfont DISCUSSION AND CONCLUSION}
\vspace{-3mm}
Presently most methods for creating disordered ensembles that are consistent with available experimental solution data are separated into two steps. The first is to create a static pool of conformations, and in a second step improving upon that pool by reweighting different sub-populations of conformations to improve a score that reflects better experimental agreement.\cite{Hummer2015,Brookes2016,Bottaro2018,Kofinger2019,Bottaro2020} If the underlying static pool is insufficient, i.e., if relevent conformations are absent, there is little that can be solved with reweighting approaches. Instead the first step needs to be revisited to create new structural pools in hope that the new underlying basis set of conformations can be made more consistent with experimental observables. Alternatively methods have been developed that generate new conformations by biasing molecular dynamics simulations with experimental data\cite{Bonomi2016,Bonomi2017}, but these methods are overall very expensive, can create significant conformational biases due to poor force fields, and/or have inadequate conformational sampling. \\

This work greatly improves upon such existing approaches by directly coupling the two steps in a machine learning method that simultaneously physically changes the conformations of the underlying pool to evolve to new structural ensembles that agree with experimental solution data at minimal computational cost and with no inherent biases. In particular the generative reinforcement ML model DynamICE biases the probability of the residue torsions of a chain molecule, generating new sub-populations of disordered states using a reward mechanism that simultaneously improves agreement with experimental data based on X-EISD scores. Currently DynamICE biases the probability distributions of torsions to take advantage of experimental data types such as J-couplings, NOEs and PREs, but extensions to other data types such as smFRET, chemical shifts and SAXS are certainly possible. \\ 

As proof-of-concept of the DynamICE method, we applied this approach by biasing toward experimental $^3$JCs and NOEs for the unfolded state of the DrkN SH3 domain and $^3$JCs and PREs for the $\alpha$-synuclein IDP. We showed that DynamICE generates ensembles of vastly different underlying structural characteristics from their starting pools, to better conform to their individual experimental data restraints. However, driving a model that uses an internal coordinate representation of protein conformers to meet distance restraints such as NOEs and PREs, is not yet fully optimal. This is a limitation of the torsion-based (local) protein representation, and not of the proposed generative-reinforcement approach. To utilize more effectively the distance/contact-based data in the reward function, we will in the future explore the use of a message passing neutral network (MPNN), which can represent the 3D coordinates of the protein conformers directly. \\

While reinforcement learning has proven its potential in \textit{de novo} drug target generation\cite{Jeon2020,Li2021} and small molecule conformer search\cite{Gogineni2020}, to our knowledge no previous attempts have been made to employ reinforcement learning to generate structural ensembles of IDPs. Posing IDP conformer generation as a RL problem introduces several benefits over the recent generative models that simply learn to represent a conformational landscape, or reweighing methods that require that all relevant conformations be present. In summary, the RL-GRNN method provides a natural framework to combine the scoring and conformer generation steps simultaneously with the experimental data, as opposed to requiring a separation of the scoring and conformer search of a starting/training conformer pool. It also greatly reduces the bias and expense of MD sampling approaches, and still considers the various errors and uncertainties of a Bayesian model. In summary, we believe that the RL-GRNN approach is a paradigm shift in how to address the overall conformational search problem for disordered states of proteins by allowing the underlying structural pools to \textit{evolve} toward experimental data under a Bayesian model that reflects statistical uncertainties, without force field bias, and includes the accounting of the Boltzmann statistics for energetically viable states.


\section*{\fontsize{12}{12}\selectfont 
AUTHOR CONTRIBUTIONS}
\vspace{-3mm}
O.Z., M. H. and T.H.-G. designed the project. O.Z. and M.H. designed and wrote the GRNN software. J.L. provided valuable comments on the design and training of the neural network model. O.Z. and T.H.-G. wrote the paper and all authors provided valuable input and discussion and editing of the final version.

\section*{\fontsize{12}{12}\selectfont ACKNOWLEDGEMENTS}
\vspace{-3mm}
All authors acknowledge funding and thank the support from the National Institute of Health under Grant 5R01GM127627-04. J.D.F.-K. also acknowledges support from the Natural Sciences and Engineering Re-search Council of Canada (2016-06718) and from the Canada Research Chairs Program.

\bibliographystyle{unsrtnat}
\bibliography{references}

\end{document}


\title{\textbf{Learning to Evolve Structural Ensembles of Unfolded and Disordered Proteins Using Experimental Solution Data}}
\author{Oufan Zhang$^{1}$, Mojtaba Haghighatlari$^{1}$, Jie Li$^{1}$, J\~{o}ao Miguel Correia Teixeira$^{3,4}$,\\
Ashley Namini$^{3}$, Zi Hao Liu$^{3,4}$, Julie D Forman-Kay$^{3,4}$, Teresa Head-Gordon$^{1,2}$\\[-5.0ex]}
\date{}
\maketitle
\noindent
\begin{center}
$^1$Kenneth S. Pitzer Theory Center and Department of Chemistry\\
$^2$Departments of Bioengineering and Chemical and Biomolecular Engineering\\
University of California, Berkeley, CA, USA
$^3$Molecular Medicine Program, Hospital for Sick Children, Toronto, Ontario M5S 1A8, Canada\\
$^4$Department of Biochemistry, University of Toronto, Toronto, Ontario M5G 1X8, Canada
\end{center}

\newpage
\section{Computational Details}

\subsection{Training procedure for the generative model} 
Separate models are trained for the unfoleded states of drkN SH3 domain and $\alpha$-synuclein. The uDrkN-SH3 pool contains 7373 conformers, and is split into 6000 for training, 600 for validation and 737 for testing. The $\alpha$-Syn pool contains 4903 conformers in total, and is split into 4000 for training, 400 for validation and rest for testing. We use categorical cross entropy loss:
\begin{align}
  L_{\Theta}=-\frac{1}{N}\sum_{i=1}^N\sum_{t_i}\hat{p}(t_i|t_1, t_2,...,t_{i-1}) \log{p_{\Theta}(t_i|t_1, t_2,...,t_{i-1})}
\end{align}
 where N represents the number of angle bins, $\hat{p}(t_i|t_1, t_2,...,t_{i-1})$ represents the actual probability of a specific torsion at the $t_i$th step, and ${p_{\Theta}(t_i|t_1, t_2,...,t_{i-1})}$ the probability predicted by the neural network with parameters $\Theta$. The model is trained using Adam optimizer\cite{kingma2014adam} in batches of size 100. To achieve convergence we employed an initial learning rate of 0.0005 and reduced the learning rate by a factor of 0.8 when the loss function plateaus. The generative models are trained for 300 epochs.
 
 \subsection{Reinforcement learning procedure}
 During RL training, torsion angles unrelated to the experimental observables being optimized are unrestrained and can lead to a noisy action space. Thus only relevant model parameters are updated while the rest remain fixed. We test JC:NOE(PRE) reward weight hyperparameters of 1:1, 1:2, and 1:4. For both RL models, a JC:NOE(PRE) reward weight of 1:4 yields the best result. In each iteration, 50 molecules are sampled, and model weights are updated by taking gradient steps on the reward function, using Adam optimizer with a learning rate cap of 0.0005. 
 
\subsubsection{Internal-Cartesian conversion}
For evaluations on the distance-based experimental data types, the conformers which are represented by torsion angle trajectories in the generative model need to be reconstructed in terms of Cartesian coordinates. We use SidechainNet package\cite{King2021} for internal to Cartesian conversion following the natural extension reference frame (NeRF) algorithm\cite{Alquraishi2019parallelized}. 

\subsection{X-EISD calculation and ensemble characterization}
The X-EISD method applies a maximum likelihood estimator to formulate a log likelihood as the degree to which a simulated ensemble is in agreement with a set of experimental data, given both the experimental and back-calculation uncertainties modeled as optimized Gaussian random variables under a Bayesian framework. X-EISD can be applied to generated an aggregated score of multiple data types as shown in Eq. \ref{eqn:eisd},
\begin{align}
    log p(X,\xi | D,I) = log p(X|I) + \sum_{j=1}^M log[p(d_j | X,\xi_j,I)p(\xi_j | I)]  + C \label{eqn:eisd}
\end{align}
where X is a set of conformers, $\xi$ denotes the various uncertainties, D is the experimental data and I is any other prior information. We use X-EISD as a probabilistic score in a simple direct maximization, performing 10,000 attempts to exchange one conformer with another for an ensemble with 100 starting structures and accepting the exchange if the new ensemble receives a higher X-EISD score than the previous. 
\begin{align}
    acc(i \rightarrow j) = X-EISD(i) > X-EISD(j) \label{eqn:MC}
\end{align}
Optimizations with each set of data type condition are repeated 100 times. In addition to the data types included during the RL training, we validate the generated ensembles with chemical shifts (CS), smFRET $\langle E\rangle$, and SAXS. We use UCBShift\cite{Li2020} for chemical shift calculations, CRYSOL software program\cite{Svergun1995} for SAXS intensities, and the efficiencies of the energy transfer are treated using in-house scripts as reported previously\cite{Lincoff2020}. The preparations of experimental data and back calculation uncertainties for the reported data types are also described previously\cite{Lincoff2020}. 

\newpage
\section{Supporting Tables}
\begin{table}[h!]
\centering
\caption{\textbf{Evaluation of the unoptimized ensembles of the original structural pool and unbiased generative model for the unfolded states of the drkN SH3 domain and $\alpha$-synuclein with experimental data types and geometric measures.} The experimental data RMSDs including J-couplings (JC), Nuclear Overhauser Effect (NOEs), Paramagnetic Relaxation Enhancement (PREs), single molecule FRET (smFRET), chemical shifts (CS), small angle X-ray scattering (SAXS). Global metrics including radius of gyration $R_g$, end-to-end distance $R_{ee}$ and asphericity $\delta^\ast$ (which measures the anisotropy of the structural ranging from 0 (sphere) to 1 (rod) are calculated using the MDTraj package\cite{Mcgibbon2015}. All values are reported in terms of mean and standard deviation(in parenthesis) over 50 ensembles of 100 structures each.}
\resizebox{\textwidth}{!}{\begin{tabular}{lccccccccc}
 & JC & NOE & PRE & smFRET $\langle E\rangle$ & CS & SAXS & $R_g$ & $R_{ee}$ & $\delta^\ast$\\
 & (Hz) & (\AA) & (\AA) & & (ppm) & (Intensity) & (\AA) & (\AA) & \\
\hline
\multicolumn{6}{l}{\textbf{UNOPTIMIZED uDrkN-SH3 with loop/extended start pool}} \\
\hline
Original pool & 1.522 & 6.590 & 7.772 & 0.222 & 0.499 & 0.007 & 22.78 & 56.47 & 0.426 \\
   & (0.036) & (0.341) & (1.289) & (0.031) & (0.009) & (0.000) &(4.37) & (21.77) & (0.188) \\ 
Generative model  & 1.440 & 6.343 & 7.711 & 0.228 & 0.495 & 0.007 & 23.16 & 55.51 & 0.431 \\
& (0.028) & (0.429) & (1.193) & (0.032) & (0.007) & (0.000) & (4.81) & (21.21) & (0.202) \\ 
\hline
\multicolumn{6}{l}{\textbf{UNOPTIMIZED $\alpha$-Syn with loop/extended start pool}} \\
\hline
Original pool & 0.709 & & 9.918 & 0.112 & 0.555 & 0.013 & 36.75 & 84.58 & 0.437 \\
 & (0.034) & & (0.428) & (0.005) & (0.004) & (0.001) & (7.91) & (31.97) & (0.193) \\ 
Generative model & 0.704 & & 10.088 & 0.108 & 0.558 & 0.013 & 37.18 & 84.73 & 0.443 \\
 & (0.022) & & (0.395) & (0.005) & (0.003) & (0.001) & (7.90) & (34.32) & (0.187) \\
\hline
\multicolumn{6}{l}{\textbf{UNOPTIMIZED $\alpha$-Syn with loop/helix start pool}} \\
\hline
Original pool & 0.583 & & 9.753 & 0.103 & 0.667 & 0.018 & 33.68 & 79.19 & 0.437 \\
 & (0.049) & & (0.434) & (0.003) & (0.026) & (0.002) & (6.60) & (28.68) & (0.194) \\ 
Generative model & 0.622 & & 9.923 & 0.103 & 0.612 & 0.017 & 33.99 & 78.66 & 0.426 \\
 & (0.032) & & (0.351) & (0.004) & (0.019) & (0.003) & (7.61) & (33.80) & (0.196) \\
\bottomrule
\end{tabular}}
\label{table:pretrain}
\end{table}

\begin{table}[h!]
\centering
\caption{\textbf{Evaluation of the RL-GRNN and reweighted ensemble optimizations for $\alpha$-Syn with experimental data types and geometric measures for ensembles optimized with JCs and PREs within 10 AA.} The experimental data RMSDs including J-couplings (JCs), nuclear overhauser effect (NOEs), paramagnetic relaxation enhancement (PREs), single molecule FRET (smFRET), chemical shifts (CS), small angle X-ray scattering (SAXS). The experimental (exp) and back calculations (back) errors for CSs ($\sigma_{exp}$=0.03-0.3 ppm); JCs ($\sigma_{exp}$=0.5); NOEs ($\sigma_{exp}$=5.0 Å); PREs ($\sigma_{exp}$=5.0 Å); smFRET <E> ($\sigma_{exp}$=0.02); SAXS ($\sigma_{exp}$=0.0008-0.002). Global metrics of the ensembles includes radius of gyration $R_g$, end-to-end distance $R_{ee}$ and ensemble asphericity $\delta^\ast$ (which measures the anisotropy of the structures ranging from 0 (sphere) to 1 (rod)). All values are reported in terms of mean and standard deviation (in parenthesis) over 50 ensembles of 100 structures each.}
\resizebox{\textwidth}{!}{\begin{tabular}{lccccccccc}
 & JC & NOE & PRE & smFRET $\langle E\rangle$ & CS & SAXS & $R_g$ & $R_{ee}$ & $\delta^\ast$\\
 & (Hz) & (\AA) & (\AA) & & (ppm) & (Intensity) & (\AA) & (\AA) & \\
\hline
\multicolumn{10}{l}{\textbf{$\alpha$-Syn ensembles UNOPTIMIZED (helix/loop) and OPTIMIZED with JCs and PREs (within 10 AA)}} \\
\hline
Generative model & 0.622 & & 7.500 & 0.103 & 0.612 & 0.017 & 33.99 & 78.66 & 0.426 \\
& (0.032) & & (0.138) & (0.004) & (0.019) & (0.002) & (7.61) & (33.80) & (0.196) \\ 
Reweight & 0.401 & & 7.109 & 0.104 & 0.594 & 0.014 & 35.30 & 81.78 & 0.438 \\
& (0.026) & & (0.144) & (0.004) & (0.018) & (0.001) & (7.46) & (30.98) & (0.195) \\ 
RL-GRNN model & 0.518 & & 6.389 & 0.106 & 0.529 & 0.016 & 38.85 & 89.95 & 0.424 \\
& (0.020) & & (0.163) & (0.006) & (0.003) & (0.002) & (7.66) & (34.56) & (0.193) \\
\hline
\multicolumn{10}{l}{\textbf{$\alpha$-Syn ensembles UNOPTIMIZED (loop/extended) and OPTIMIZED with JCs and PREs (within 10 AA)}} \\
\hline
Generative model & 0.704 & & 7.225 & 0.108 & 0.558 & 0.013 & 37.18 & 84.73 & 0.443 \\
 & (0.022) & & (0.157) & (0.005) & (0.003) & (0.001) & (7.90) & (34.32) & (0.187) \\ 
Reweight & 0.444 & & 6.976 & 0.108 & 0.543 & 0.013 & 35.30 & 81.78 & 0.438 \\
 & (0.034) & & (0.153) & (0.006) & (0.007) & (0.001) & (7.46) & (30.98) & (0.195) \\ 
RL-GRNN model & 0.593 & & 6.335 & 0.114 & 0.570 & 0.014 & 37.79 & 86.19 & 0.406 \\
 & (0.011) & & (0.157) & (0.006) & (0.002) & (0.001) & (7.78) & (31.75) & (0.186) \\
\hline
\bottomrule
\end{tabular}}
\label{table:asyn_LE}
\end{table}

\clearpage
\section{Supporting Figures}

\begin{figure}[ht!]
\centering
\includegraphics[width=17cm]{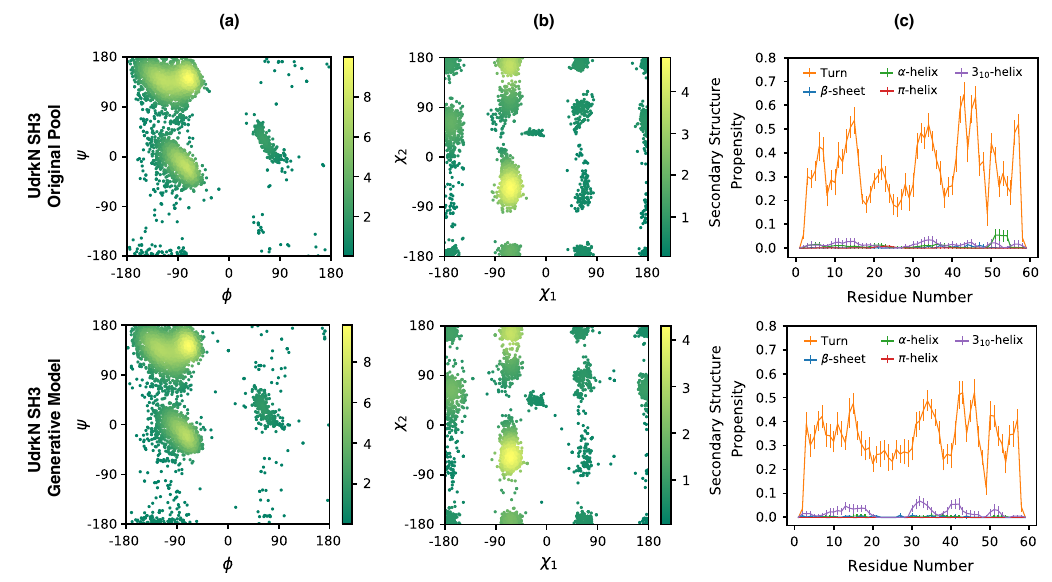}
\caption{\textbf{Properties of ensembles for uDrkN-SH3 from the original pool and from the generative model.} a) Ramachandran plots displaying the backbone torsion angle distributions and b) Histograms displaying the $\chi_1-\chi_2$ distributions from 100 structures of the training data (top) and generative model (bottom). Density values are scaled by 1e-05. c). Secondary structure propensities per residue among 50 independently drawn ensembles of 100 structures. Error bars are shown as $\pm$ 1 standard deviation.}
\label{fig:drk_pretrain}
\end{figure}

\begin{figure}[ht!]
\centering
\includegraphics[width=17cm]{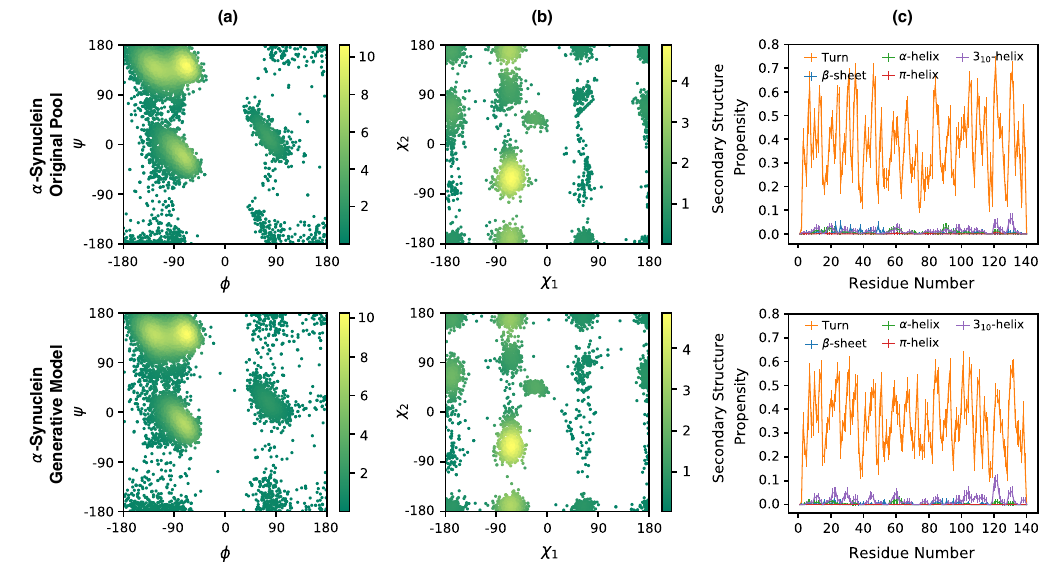}
\caption{\textbf{Properties of ensembles for $\alpha$-Syn from the loop/extended original pool and from the generative model.} a) Ramachandran plots displaying the backbone torsion angle distributions and b) Histograms displaying the $\chi_1-\chi_2$ distributions from 100 structures of the training data (top) and generative model (bottom). Density values are scaled by 1e-05. c). Secondary structure propensities per residue among 50 independently drawn ensembles of 100 structures. Error bars are shown as $\pm$ 1 standard deviation.}
\label{fig:asyn_pretrain}
\end{figure}

\begin{figure}[!ht]
\centering
\includegraphics[width=17cm]{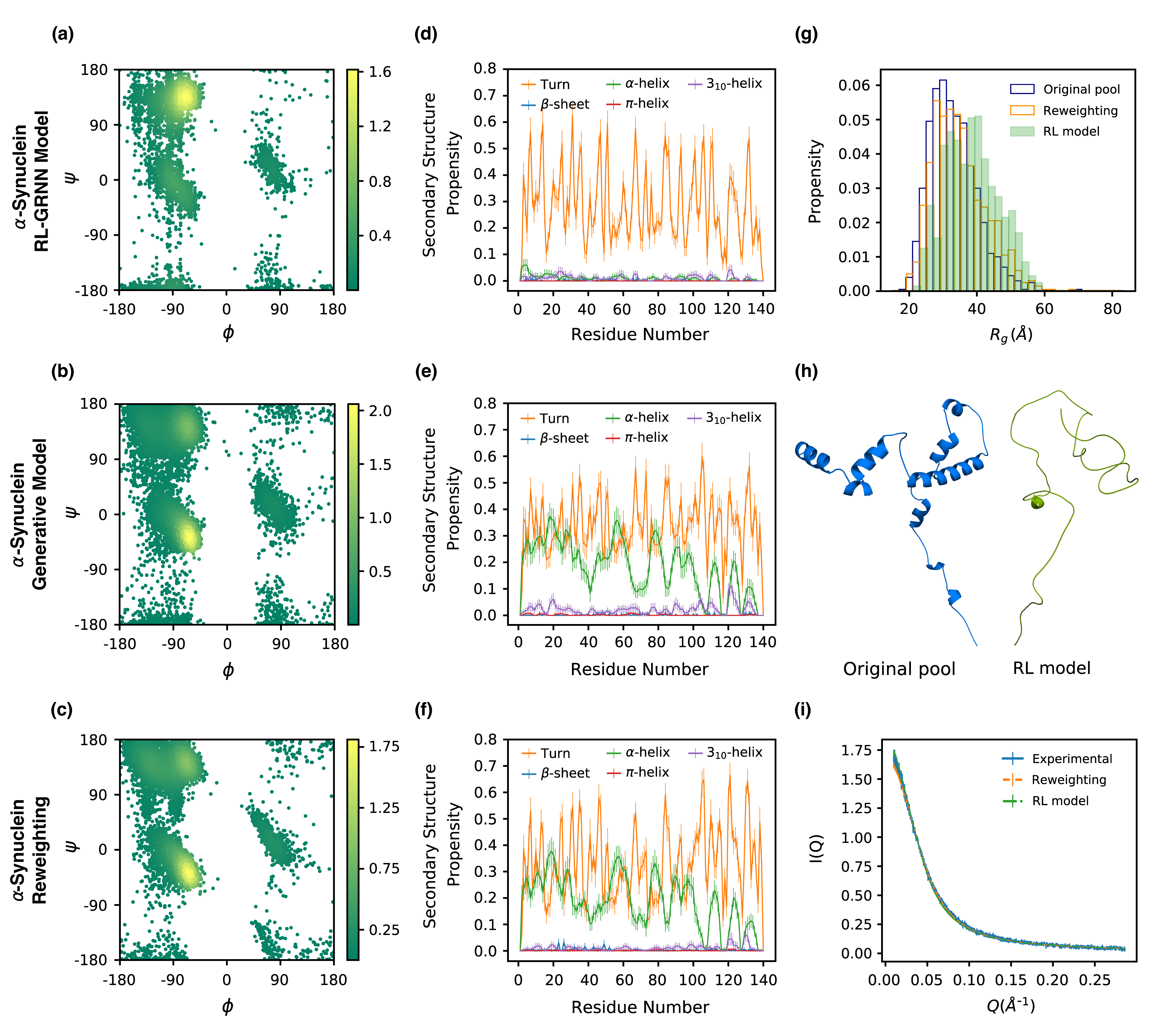}
\caption{\textbf{Properties of the $\alpha$-Syn unbiased ensemble of loops/helix states compared to optimized ensembles using JCs and PREs within 10 AA generated by the DynamICE RL-GRNN model compared with reweighting optimization.} Ramachandran plots displaying the backbone torsion angle distributions from the (a) RL-GRNN, (b) unbiased generative model, and (c) reweighting optimization. Density values are scaled by 1e-04. Secondary structure propensities per residue of the (d) RL-GRNN, (e) unbiased generative  model, (f) reweighting optimization. (g) Comparison of radius of gyration distributions before and after optimization with reweighting optimization and RL-GRNN. (h) Example of conformers from the $\alpha$-Syn original pool of loops/helix states and RL-GRNN model. (i) SAXS intensity curves for RL-GRNN and reweighting optimized ensembles compared with the experimental data. SAXS intensity is scaled by 0.001. Statistical errors from 50 independently drawn ensembles of 100 structures. Error bars are shown as $\pm$ 1 standard deviation.}
\label{fig:asyn_rl}
\end{figure}



\clearpage
\bibliographystyle{unsrtnat}
\bibliography{references}